\documentclass{procmult}
\usepackage{graphicx}
\usepackage{amsmath}
\usepackage{amsfonts}

\usepackage{amssymb}

\begin{document}

\title{Stochastic background of gravitational waves "tuned" by $f(R)$ gravity}
\author{Mariafelicia De Laurentis$^\dag$, Salvatore Capozziello$^\dag$ \and Luca Izzo$^\dag$ $^\ddag$}
\institute{$^\dag$ Dipartimento di Scienze Fisiche, Universit\`a di
Napoli "Federico II" \\and INFN, Sez. di Napoli, Italy\\
$^\ddag$  ICRANet and ICRA,  Pescara, Italy, and Dipartimento di Fisica, Universit\`a di Roma "La Sapienza", Roma, Italy.}
\maketitle
\abstract{Abstract}
The cosmological background of gravitational waves can be tuned by
Extended Theories of Gravity. In particular, it can be shown that
assuming a generic function $f(R)$ of the Ricci scalar $R$  gives
a parametric approach to control the evolution and the production
mechanism of gravitational waves in the early Universe.

\section{Introduction}
 In the last thirty years several shortcomings
came out in  Einstein General Relativity (GR) and people began to
investigate whether it is the only theory capable of explaining
gravitational interactions. Such issues  sprang up in Cosmology
and Quantum Field Theory. In the first case, the  Big Bang
singularity, the flatness and horizon problems led to the result
that the Standard Cosmological Model is inadequate to describe the
Universe in extreme regimes. Besides, GR is a \textit{classical}
theory which does not work as a fundamental theory. Due to these
facts and to the lack of a self-consistent Quantum Gravity theory,
alternative theories have been pursued.  A fruitful approach is
that of Extended Theories of Gravity (ETGs) which have become a
sort of paradigm based on corrections and enlargements of GR.

These theories have acquired interest in Cosmology owing to the
fact that they ``naturally" exhibit inflationary behaviors
\cite{starobinsky}. Recently, ETGs are
 playing  an interesting role in describing
  the today observed
Universe. In fact, the amount of good quality data of last decade
has made it possible to shed new light on the effective picture of
the Universe.  In particular, the \textit{Concordance  Model}
predicts that baryons contribute only for $\sim4\%$ of the total
matter\,-\,energy budget, while the exotic \textit{cold dark
matter}  represents the bulk of the matter content ($\sim25\%$)
and the cosmological constant $\Lambda$ plays the role of the so
called "dark energy" ($\sim70\%$). Although being the best fit to
a wide range of data, the $\Lambda$CDM model is severely affected
by strong theoretical shortcomings that have motivated the search
for alternative models. Dark energy models mainly rely on the
implicit assumption that GR is the correct theory of gravity
indeed. Nevertheless, its validity on the larger astrophysical and
cosmological scales has never been tested, and it is therefore
conceivable that both cosmic speed up and dark matter represent
signals of a breakdown in GR
\cite{frafe,cap,odintsov1,allemandi,odintsov2,GRGrev,faraoni}.

From an astrophysical viewpoint, ETGs do not require  to find out
candidates for dark energy and dark matter at  fundamental level;
the approach starts from taking into account only the ``observed''
ingredients (i.e. gravity, radiation and baryonic matter); this is
in agreement with the early spirit of GR which could not act in
the same way at all scales. In fact, GR has been definitively
probed in the weak field limit and up to Solar System scales.
However, a comprehensive effective theory of gravity, acting
consistently at any scale, is far, up to now, to be found, and
this demands an improvement of observational datasets and the
search for experimentally testable theories. A  pragmatic point of
view could be to ``reconstruct'' the suitable theory of gravity
starting from data. The main issues of this ``inverse '' approach
is matching consistently observations at different scales and
taking into account wide classes of gravitational theories where
``ad hoc'' hypotheses are avoided. In principle, the most popular
dark energy models can be achieved by considering  $f(R)$ theories
of gravity
 and the same track can be followed to match galactic dynamics \cite{CCT}.
 This philosophy can be
taken into account also for the cosmological stochastic background
of gravitational waves (GW) which, together with CMBR, would
carry, if detected, a huge amount of information on the early
stages of the Universe evolution.

\section{Gravitational waves in f(R)-gravity}
In this paper,  we  face the problem to match  a generic $f(R)$
theory with the cosmological background of GWs. GWs are
perturbations $h_{\mu\nu}$ of the metric $g_{\mu\nu}$ which
transform as 3-tensors.  The GW-equations in the
transverse-traceless gauge are
\begin{equation}
\square h_{i}^{j}=0\label{eq: 1},
\end{equation}
  Latin indexes run from 1
to 3. Our task is now to derive the analog of Eqs.\ (\ref{eq: 1})
from a generic $f(R)$ given by the action
\begin{equation}
\mathcal{A}=\frac{1}{2k}\int d^{4}x\sqrt{-g}f(R)\label{eq:2}\,.
\end{equation}
From conformal transformation, the extra degrees of
 freedom related to  higher order gravity can be recast into a
 scalar field
\begin{equation}
\widetilde{g}_{\mu\nu}=e^{2\Phi}g_{\mu\nu}\qquad \mbox{with}
\qquad e^{2\Phi}=f'(R)\,.\label{eq:3}
\end{equation}
Prime indicates the derivative with respect to  $R$. The
conformally equivalent Hilbert-Einstein action is
\begin{equation}
\mathcal{\widetilde{A}}=\frac{1}{2k}\int\sqrt{-\tilde{g}}d^{4}x\left[\widetilde{R}+
\mathcal{L}\left(\Phi\mbox{,}\Phi_{\mbox{;}\mu}\right)\right]\label{eq:4}\end{equation}
where $\mathcal{L}\left(\Phi\mbox{,}\Phi_{\mbox{;}\mu}\right)$ is
the  scalar field Lagrangian derived from
\begin{equation}
\widetilde{R}=e^{-2\Phi}\left(R-6\square\Phi-6\Phi_{;\delta}\Phi^{;\delta}\right)\,.\label{eq:6}\end{equation}
 The GW-equation is now
\begin{equation}
\widetilde{\square}\tilde{h}_{i}^{j}=0\label{eq:7}
\end{equation}
where
\begin{equation}
\widetilde{\square}=e^{-2\Phi}\left(\square+2\Phi^{;\lambda}\partial_{;\lambda}\right)\label{eq:9}\,.\end{equation}
Since no scalar perturbation couples to the tensor part of
gravitational waves, we have
\begin{equation}
\widetilde{h}_{i}^{j}=\widetilde{g}^{lj}\delta\widetilde{g}_{il}=e^{-2\Phi}g^{lj}e^{2\Phi}\delta
g_{il}=h_{i}^{j}\label{eq:8}
\end{equation}
which means that $h_{i}^{j}$ is a conformal invariant.

As a consequence, the plane-wave amplitudes
$h_{i}^{j}(t)=h(t)e_{i}^{j}\exp(ik_{m}x^{m}),$ where $e_{i}^{j}$
is the polarization tensor, are the same in both metrics. This
fact will assume a key role in the following discussion.
\section{Cosmological stochastic background of gravitational waves}
In a FRW background,  Eq.(\ref{eq:7}) becomes
\begin{equation}
\ddot{h}+\left(3H+2\dot{\Phi}\right)\dot{h}+k^{2}a^{-2}h=0\label{eq:10}
\end{equation}
being  $a(t)$ the scale factor,  $k$ the wave number and $h$ the
GW amplitude. Solutions are combinations of Bessel's functions.
Several mechanisms can be considered for the production of
cosmological GWs.  In principle, we could seek for contributions
due to every high-energy  process in the early phases of the
Universe.

In the case of inflation, GW-stochastic background is strictly
related to dynamics of cosmological model. This is the case we are
considering here. In particular, one can assume that the main
contribution to the stochastic background comes from the
amplification of vacuum fluctuations at the transition between the
inflationary phase and the radiation  era. However,  we can assume
that the GWs generated as zero-point fluctuations during the
inflation undergo adiabatically damped oscillations $(\sim 1/a)$
until they reach the Hubble radius $H^{-1}$. This is the particle
horizon for the growth of perturbations. Besides, any previous
fluctuation is smoothed away by the inflationary expansion. The
GWs freeze out for $a/k\gg H^{-1}$ and reenter the $H^{-1}$ radius
after the reheating. The reenter in the Friedmann era depends on
the scale of the GW. After the reenter, GWs can be detected by
their Sachs-Wolfe effect on the temperature anisotropy
$\bigtriangleup T/T$ at the decoupling. If $\Phi$ acts as the
inflaton, we have $\dot{\Phi}\ll H$ during the inflation. Adopting
the conformal time $d\eta=dt/a$, Eq.\ (\ref{eq:10}) reads
\begin{equation}
h''+2\frac{\chi'}{\chi}h'+k^{2}h=0\label{eq:16}
\end{equation}
where $\chi=ae^{\Phi}$. The derivation is  now with respect to
$\eta$.  Inside the  radius $H^{-1}$, we have $k\eta\gg 1.$
Considering the absence of gravitons in the initial vacuum state,
we have only negative-frequency modes and then the solution of
(\ref{eq:16}) is
\begin{equation}
h=k^{1/2}\sqrt{2/\pi}\frac{1}{aH}C\exp(-ik\eta)\,.\label{eq:18}
\end{equation}
 $C$ is the amplitude parameter. At the first horizon crossing
$(aH=k)$ the averaged amplitude
$A_{h}=(k/2\pi)^{3/2}\left|h\right|$ of the perturbation is
\begin{equation}
A_{h}=\frac{1}{2\pi^{2}}C\,.\label{eq:19}
\end{equation}
When the scale $a/k$ becomes larger than the Hubble radius
$H^{-1}$, the growing  mode of evolution is constant, i.e. it is
frozen.   It can be shown that $\bigtriangleup T/T\lesssim A_{h}$,
as an upper limit to $A_{h}$, since other effects can contribute
to the background anisotropy. From this consideration, it is clear
that the only relevant quantity is the initial amplitude $C$ in
Eq.\ (\ref{eq:18}), which is conserved until the reenter. Such an
amplitude depends  on the fundamental mechanism generating
perturbations. Inflation gives rise to processes capable of
producing perturbations as zero-point energy fluctuations. Such a
mechanism depends on the gravitational interaction and then
$(\bigtriangleup T/T)$ could constitute a further constraint to
select a suitable theory of gravity. Considering a single graviton
in the form of a monochromatic wave, its zero-point amplitude is
derived through the commutation relations:
\begin{equation}
\left[h(t,x),\,\pi_{h}(t,y)\right]=i\delta^{3}(x-y)\label{eq:20}
\end{equation}
calculated at a fixed time $t$, where the amplitude $h$ is the
field and $\pi_{h}$ is the conjugate momentum operator. Writing
the Lagrangian for $h$
\begin{equation}
\widetilde{\mathcal{L}}=\frac{1}{2}\sqrt{-\widetilde{g}}\widetilde{g}^{\mu\nu}h_{;\mu}h{}_{;\nu}\label{eq:21}
\end{equation}
in the conformal FRW metric $\widetilde{g}_{\mu\nu}$, where the
amplitude $h$ is conformally invariant, we obtain
\begin{equation}
\pi_{h}=\frac{\partial\widetilde{\mathcal{L}}}{\partial\dot{h}}=e^{2\Phi}a^{3}\dot{h}\label{eq:22}\end{equation}

Eq.\ (\ref{eq:20}) becomes
\begin{equation}
\left[h(t,x),\,\dot{h}(y,y)\right]=i\frac{\delta^{3}(x-y)}{a^{3}e^{2\Phi}}\label{eq:23}
\end{equation}
and the fields $h$ and $\dot{h}$ can be expanded in terms of
creation and annihilation operators. The commutation relations in
conformal time are
\begin{equation}
\left[hh'^{*}-h^{*}h'\right]=\frac{i(2\pi)^{3}}{a^{3}e^{2\Phi}}\,.\label{eq:26}
\end{equation}
From (\ref{eq:18}) and (\ref{eq:19}), we obtain
$C=\sqrt{2}\pi^{2}He^{-\Phi}$, where $H$ and $\Phi$ are calculated
at the first horizon-crossing and, being $e^{2\Phi}=f'(R)$, the
relation
\begin{equation}
A_{h}=\frac{H}{\sqrt{2f'(R)}}\,, \label{eq:27}
\end{equation}
holds for a generic $f(R)$ theory. 

\section{Conclusion}
The central result (eq. (\ref{eq:27})) of
this paper deserves some discussion. Clearly the amplitude of
GWs produced during inflation depends on the theory of gravity
which, if different from GR, gives extra degrees of freedom.  On
the other hand, the Sachs-Wolfe effect could constitute a test for
gravity at early epochs. This probe could give further constraints
on the GW-stochastic background, if ETGs are independently probed
at other scales.

\end{document}